\documentstyle[aps,epsf]{revtex}
\input epsf.sty
\def\ua{\uparrow}
\def\da{\downarrow}
\def\prb{Phys. Rev. B}
\def\prl{Phys. Rev. Lett.}

\def\be{\begin{equation}}
\def\ee{\end{equation}}
\def\ba{\begin{eqnarray}}
\def\ea{\end{eqnarray}}

\def\C60{A$_x$C$_{60}$}

\def\hty{high temperature superconductivity}
\def\hts{high temperature superconductors}
\begin{document}
\draft
\flushbottom
\twocolumn[
\hsize\textwidth\columnwidth\hsize\csname @twocolumnfalse\endcsname

\title{Crossovers and Phase Coherence in Cuprate Superconductors}
\author{V.~J.~Emery$^1$ and S.~A.~Kivelson$^2$}
\address{
$^1$Dept. of Physics
Brookhaven National Laboratory
Upton, NY  11973}
\address{
$^2$Dept. of Physics
University of California at Los Angeles
Los Angeles, CA 90095}
\date{\today}
\maketitle
\tightenlines
\widetext
\advance\leftskip by 57pt
\advance\rightskip by 57pt

\begin{abstract}
High temperature superconductivity is a property of 
doped antiferromagnetic insulators. The electronic structure is inhomogeneous 
on short length and time scales, and, as the temperature decreases, it 
evolves via two crossovers, before long range superconducting order is 
achieved. Except for overdoped materials, pairing and phase coherence occur 
at different temperatures, and phase fluctuations determine both T$_c$ and 
the temperature dependence of the superfluid density for a wide range of 
doping. A mechanism for obtaining a high pairing scale in a short coherence
length material with a strong poorly-screened Coulomb interaction 
is described.

\end{abstract}

]

\narrowtext
\tightenlines

\section{Introduction}

High temperature superconductivity \cite{BM} is a property of quasi-two 
dimensional doped insulators, obtained by chemically introducing charge 
carriers into a highly-correlated antiferromagnetic insulating state. 
There is a large ``Fermi surface'' containing all of the 
holes in the relevant Cu(3d) and O(2p) orbitals \cite{arpes}, but $n/m^*$ 
vanishes as the dopant concentration tends to zero.\cite{muon,optical} 
(Here $m^*$ is the effective mass of a hole and $n$ is either the superfluid 
density or the density of mobile charges in the normal state.) 
The phase diagram, Fig. 1, shows that 
superconductivity occurs in a narrow range of doping close to the 
antiferromagnetic insulating state, and emerges
gradually as the temperature is lowered, via two {\it crossovers} at which 
specific {\it local} electronic structure develops,\cite{batloggemery,ekz} 
and a {\it phase transition} where {\it long-range} phase order is established. 
Clearly, understanding the origin of {\hty} and the nature of the doped 
insulating state go hand in hand.

In our view, the driving force behind all of this behavior is the tendency
of the antiferromagnet to expel the doped holes and so to form hole rich and 
hole free regions.\cite{ekl} For neutral holes this leads to a 
{\it first-order phase transition}
(phase separation)\cite{ekl,hellberg} but, for charged holes, the competition 
with the long-range part of the Coulomb interaction generates a 
dynamical {\it local} charge inhomogeneity, in which the mobile holes are 
typically confined in charged ``stripes'', \cite{ute,spherical,zaanen}
which are antiphase domain walls for the spins in the intervening undoped 
regions. {\it Locally}, the  electronic structure has a quasi 
one-dimensional character. 
There is extensive evidence, both direct \cite{jtran,mook,cheong} and indirect
\cite{physica}, for this interpretation of the experiments.

Charge instabilities are a general consequence of the competition between
phase separation and the long-range Coulomb interaction, and they are a 
common feature of oxides in general.\cite{jtran,cheong} However, the 
the mechanism of superconductivity and the nature of the superconducting 
state depend on the details of the underlying microscopic model. Here, we are 
interested in systems with purely repulsive interactions. Models with 
an effective short-range attraction are described by Di Castro \cite{dicastro}. 

\begin{figure}
\vspace{.8cm}
\hspace{1.15 in}
\epsfysize=3.5in
\epsfbox{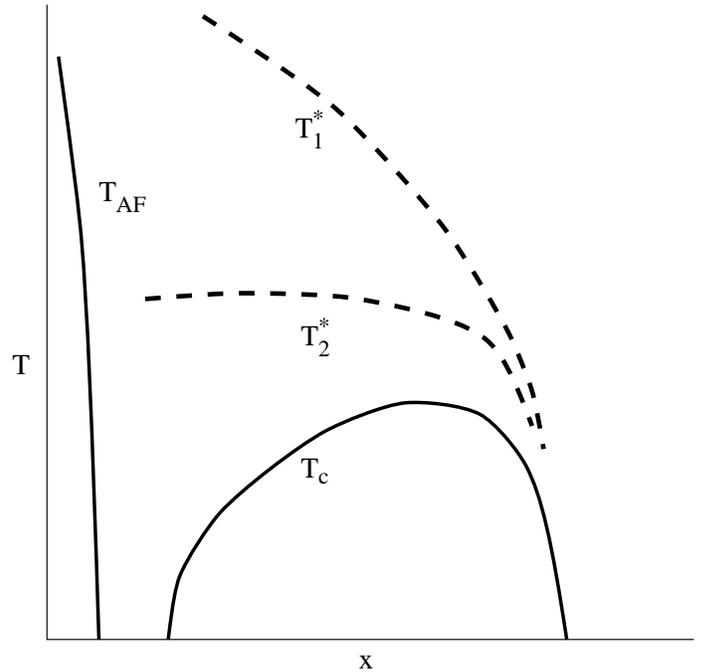}
\vspace{.8cm}
\caption{
Sketch of the
phase diagram for a high temperature superconductor in the
doping-temperature plane.  The solid lines represent phase transitions
and the broken lines indicate crossovers.  $T_N$ marks the transition to
an antiferromagnetically ordered insulating state, and $T_c$ the
transition to the superconducting state.  $T_1^*$ and $T_2^*$ mark the 
two crossovers discussed in the text.}
\label{fig1}
\end{figure}

\section{Crossovers}

A crossover signifies a change in the short- or intermediate-distance 
behavior of a system as the temperature or some other thermodynamic quantity
is varied. Unlike a phase transition, at which long range order is
established, a crossover is not abrupt, and usually it appears at slightly
different temperatures in different physical properties. 
The  existence of two crossovers in the {\hts} is evident
in NMR experiments; the Knight shift begins to decrease at a temperature 
T$_1^*$, whereas $(T_1T)^{-1}$ (where $T_1$ is the nuclear spin relaxation 
rate) does not start to decrease until a lower temperature T$_2^*$.
In underdoped materials these two temperatures are well separated. For
example, in  HgBa$_2$Ca$_2$Cu$_3$O$_{8+\delta}$, \cite{julien} T$_1^*$ is about 
370K, whereas T$_2^*$ is about 230K. 
On the other hand, as shown in Fig. 1, 
the two crossovers merge just above T$_c$ in most optimally-doped materials. 
This is why it appeared at first that there was just one crossover at a  
``spin gap '' temperature T$^*$. However, although
a drop in $(T_1T)^{-1}$ (which depends on the imaginary part of the 
spin susceptibility $\chi$) indicates the opening of a spin gap (or pseudogap),
a drop in the Knight shift (which depends on the real part of $\chi$)
could indicate either the opening of a spin gap or 
the development of short-range antiferromagnetic correlations, or both.
As  $x\rightarrow 0$, $T^*_1$ approaches the temperature at which local 
antiferromagnetic correlations, {\it not a spin gap}, develop in the 
undoped systems\cite{batlogg}. 
At finite doping, other information is necessary 
to distinguish between the different possibilities.

In underdoped materials, the $c$-axis optical conductivity $\sigma_c(\omega)$ 
develops a pseudogap at a temperature that correlates well with the upper 
crossover T$_1^*$.\cite{homes} The spectral weight is transferred upwards to 
quite high frequencies, which indicates the development of short-range
charge and/or spin correlations. On the other hand, at a lower temperature
T$_2^*$, the optical conductivity $\sigma_{ab}(\omega)$ in the $ab$-plane
develops a pseudo-delta function
or, in other words,  a narrowing of the central
``Drude-like'' peak.\cite{basov} 
Essentially all of the spectral weight from a pseudogap region moves downwards, 
which indicates the development of superconducting correlations. Other 
experiments support this general picture. In particular, angle-resolved 
photoemission spectroscopy (ARPES) shows that the pseudogap in the normal 
state has essentially the same magnitude and momentum dependence as the
gap in the supeconducting state.\cite{arpesgap} 

\subsection{Lower Crossover: Phase Fluctuations}

The existence of local superconducting correlations below $T^*_2$ indicates
that the amplitude of the order parameter is well established but there is no 
long-range phase coherence. This behavior, which may be deduced
from the experiments, regardless of the underlying microscopic model,
\cite{nature} is not unusual in the statistical mechanics of systems with a 
two-component order parameter, but it constitutes a major difference
between {\hts} and conventional superconductors, for which pairing 
and phase coherence are established at one and the same temperature. 

The pairing scale is related to the size of the coherence length $\xi_0$
or equivalently the energy gap $\Delta_0$ at zero temperature, and, in the 
BCS mean field theory, $\Delta_0/2$ provides a good estimate of T$_c$.
At the same time, the classical phase ordering temperature is 
determined by the ``phase stiffness'' $V_0$ which sets the energy scale for 
the spatial variation of the superconducting phase. The classical phase 
Hamiltonian is
\begin{equation}
H = V_0 \sum_{i,j} \cos(\theta_i - \theta_j)
\end{equation}
and, if $V_0$ is independent of $T$, the phase ordering temperature 
T$_{\theta} = AV_0$, where $A$ is a number of  order unity.\cite{nature}
At zero temperature, $V_0$ is given in terms of the superfluid density
$n_s(T=0)$ or, equivalently, the experimentally measured penetration depth 
$\lambda(T=0)$:
\begin{equation}
V_0 = {\hbar^2 n_s(0)a \over 4m^*} 
=   {(\hbar c)^2 a \over 16\pi(e\lambda(0))^2}
\end{equation}
where $a$ is a length scale that depends on the dimensionality of the material.
An estimate for T$_c$ is given by the smaller of $\Delta_0 /2$ and $T_{\theta}$.

The separation of the temperatures for pairing and phase coherence as the
doping $x$ is decreased below its optimal value could, in principle, be 
accomplished either by decreasing $\xi_0$ (increasing $\Delta_0$) and 
elevating the pairing scale, or by decreasing $n_s(0)$ and depressing the 
phase coherence scale. Figure 1 clearly shows that the separation of scales
is caused by the drop in the superfluid density (a property of a
doped insulator) and {\it not} by a decrease in $\xi_0$ 
(a crossover to 
Bose-Einstein condensation).

For conventional materials, the value of $\Delta_0/2$ gives a very good 
estimate for T$_c$ whereas, for {\it e.g.} Pb, T$_{\theta} =AV_0$ is about 
$10^6$K. \cite{nature} This is why BCS theory works so well. On the other 
hand, for underdoped {\hts}, $\Delta_0/2$ 
is closer to $T^*_2$ than to T$_c$, \cite{oda} whereas
T$_{\theta}$ is very close to T$_c$ itself, especially for underdoped 
materials. \cite{nature} In other words, because the {\hts} are
doped insulators, $n_s(0) \rightarrow 0$ as $x \rightarrow 0$, 
and phase ordering controls T$_c$. \cite{nature} 

Phase fluctuations also give a good description of the temperature
dependence of the superfluid density below T$_c$. It has been shown by 
University of British Columbia group\cite{hardy} that, if the measured values
of $\lambda^2(0)/\lambda^2(T)$ for YBa$_2$Cu$_3$O$_{7-\delta}$ with
$\delta$ equal to 0.01, 0.05, and 0.40, are plotted as functions of $T/T_c$,
they all lie on the same curve. In other words, T$_c$ is the one and only 
energy scale involved in  the temperature dependence of $\lambda(T)$ for overdoped, 
optimally doped, and underdoped samples of YBCO. Moreover, near T$_c$, all 
three samples display the critical behavior expected for 
classical phase fluctuations.
Therefore, on empirical grounds, it is difficult to escape the conclusion 
that the entire temperature dependence of $\lambda(T)$ is governed by
classical phase fluctuations. We have shown \cite{quantum} that,
in the superconducting state, phase fluctuations are indeed classical over 
a very wide temperature range because the low-frequency conductivity that 
exists in addition to the $\delta(\omega)$ peak of the superconducting 
condensate \cite{dimitri} is sufficient to screen the Coulomb interaction 
down to very low temperatures. Figure 2 shows
the temperature dependence of $\lambda^2(0)/\lambda^2(T)$ for the 
three-dimensional version of the simple classical phase Hamiltonian given in 
Eq. (1), together with a comparison with the experimental data.\cite{figure} 
Of course, in YBCO, $V_0$ should be anisotropic within the CuO$_2$ planes and
should be quite small in the direction perpendicular to the planes, so 
a more accurate model would have two additional dimensionless parameters that 
could be adjusted to fit the experiments. 
Nevertheless, it can be seen that the calculated and experimental curves are
already very close, without any tuning of parameters.
                                                    
\begin{figure}
\vspace{.8cm}
\hspace{1.15 in}
\epsfysize=3.5in
\epsfbox{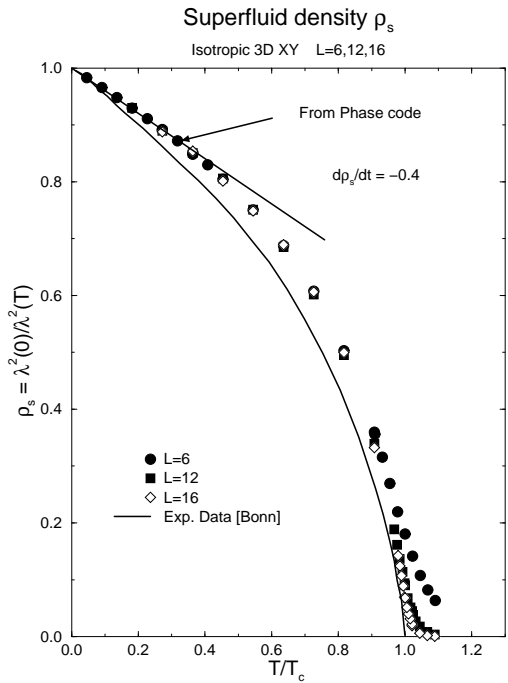}
\vspace{.8cm}
\caption{Scaling plot of the superfluid density
as a function of T/T$_c$. The theoretical curve is obtained in 
ref. \protect 
\cite{figure} from the Hamiltonian in Eq. (1), and the data are from 
ref. \protect \cite{hardy}.
}
\label{fig2}
\end{figure}

Of course this global picture is not sufficient to evaluate the effects of 
phase fluctuations on all physical properties. For example, in calculating 
the magnetoresistance, it is important to take into account the local 
electronic structure (subsec. B), 
which profoundly influences the motion of the mobile 
holes in a magnetic field.

It has been suggested that the T-linear variation of 
$\lambda^2(0)/\lambda^2(T)$ at low temperatures is a consequence of 
quasiparticle excitations near the nodes of a $d$-wave gap.
\cite{quasi,lee,millis} However the energy scale for this 
process is set by the slope of the gap at the node, and this quantity is not 
simply related to T$_c$, which is the energy scale of the experiments. 
Various ways of addressing this difficulty have been suggested 
\cite{lee,millis} but they do not have the simplicity and naturalness of 
the explanation in terms of phase fluctuations, which anyway
do not leave much room for other contributions. Moreover, quite extensive
ARPES experiments \cite{ding} have not found any evidence of quasiparticle
excitations near the nodes of the gap even just below T$_c$. 
The present data suggest that the number of quasiparticle excitations 
in the superconducting state (which should be 
proportional to the dopant concentration in a doped insulator
\cite{stripesqp}) is so small that their effect on the ARPES 
experiments and penetration depth measurements is difficult to observe.

The significance of phase fluctuations has been questioned by Geshkenbein 
{\it et al.} \cite{gil} and by Millis {\it et al.}.\cite{millis} However, in 
estimating T$_{\theta}$, these papers incorrectly assumed that $V_0$ is the 
{\it renormalized} phase coupling, whereas it is, of course, the {\it bare} 
phase coupling that appears in Eq. (1).
They made the unrealistic assumption that, in YBa$_2$Cu$_3$O$_{7-\delta}$ and 
YBa$_2$Cu$_3$O$_8$, there is a strong {\it bare} phase 
coupling between the bilayers, and made the incorrect assertion that 
bilayer coupling would double the estimate of T$_{\theta}$. 
Moreover they focussed solely on highly anisotropic materials with bilayers, 
and thereby failed to appreciate the overall picture which shows that T$_c$
is controlled by phase fluctuations in a wide variety of clean underdoped 
cuprate superconductors that do not suffer from these complications.

\subsection{Upper Crossover: Local Inhomogeneity}

The upward movement of spectral weight in $\sigma_c(\omega)$ at $T_1^*$
signifies the development of the charge and spin correlations associated with 
the formation of stripes. Locally, an individual stripe may be regarded 
as a one dimensional electron gas (1DEG)
in an active environment and, for repulsive interactions, the dominant 
instability is to the formation of charge density waves (CDW).\cite{ekz} 
However the posibility that an array of stripes might form an ordered 
insulating CDW state at low temperatures is entirely eliminated if 
the zero-point energy of transverse stripe fluctuations is sufficiently large 
in comparison to the coupling between stripes.\cite{kfe}
As a consequence, there
exist novel, liquid-crystalline low-temperature phases -- an electron smectic, 
with crystalline order in one direction, but liquid-like correlations in 
the other, and an electron nematic with orientational order but no long-range
positional order.\cite{kfe}
In the presence of symmetry breaking fields there is a crossover to the nematic 
region, rather than a phase transition. 
                     
The isotropic-to-nematic boundary has many of the characteristics of 
the upper crossover. 
At high temperatures the holes are more or less 
uniformly distributed, and randomly disrupt antiferromagnetic correlations.
However, a self-organized stripe array, especially in a nematic phase, allows 
a mixture of local antiferromagnetic correlations and spin singlets to 
develop in the hole-free regions of the sample. \cite{ekz}
Stripes tend to separate the spins into regions that are more or less
disconnected from some of their neighbors, and there is much 
numerical and analytical evidence to show that some of the low-energy spin 
degrees of freedom acquire an energy gap in such a structure.\cite{djs,ekz}
This gap is a consequence of local physics, not impending long range
antiferromagnetic order and, for this reason, it has the potential to be 
the source of superconducting pairing, as we shall see. Taking all of these
effects together, the upper crossover is signified by a drop in the magnetic 
susceptibility as well as a spontaneous breakdown of fourfold rotational 
symmetry of the CuO$_2$ planes (wherever it exists). 

As the concentration of holes increases, the separation between stripes 
eventually becomes comparable to their width and all information 
concerning the Mott insulating state is lost. Here, the isotropic-to-nematic 
line ends at a zero temperature quantum critical point. Different versions 
of such a point, either 0+1 dimensional \cite{kondo,physica} or 2+1 
dimensional \cite{dicastro,varma} have
been invoked to explain the unusual normal state properties of 
the high temperature superconductors.

\section{Pairing Mechanism}

A major problem for any mechanism of {\hty} is how to achieve a high pairing 
scale in the presence of the repulsive Coulomb interaction, especially in
a doped Mott insulator in which there is poor screening. 
In the {\hts}, the coherence length is no
more than a few lattice spacings, so neither retardation, nor a
long-range attractive interaction is effective in overcoming the bare 
Coulomb repulsion. Nevertheless ARPES experiments \cite{darpes} show that 
the major component of the energy gap is
$\cos k_x - \cos k_y$. Since the Fourier transform of this quantity vanishes
unless the distance is one lattice spacing, it follows that
the gap (and hence, in BCS theory, the net pairing force) is a maximum for holes 
separated by one lattice spacing, where the bare Coulomb interaction is very 
large ($\sim$ 0.5 eV, allowing for atomic polarization). It is not easy to 
find a source of an attraction that is strong enough to overcome such a Coulomb
force at short distances {\it and} 
achieve {\hty} via the usual Cooper pairing. 

\subsection{Spin Gap Proximity Effect}

The stripe structure provides a very natural way to overcome this problem.
For a 1DEG, the singlet pair operator may be written
\begin{equation}
\psi^{\dagger}_{1 \ua} \psi^{\dagger}_{2 \da}  +
\psi^{\dagger}_{2 \ua} \psi^{\dagger}_{1 \da} \sim e^{i\theta_c} \cos \phi_s                        
\end{equation}
Here $\phi(x)$ is a Bose field and $\pi(x)$ its conjugate momentum,
$\partial_x \theta \equiv \pi$ and the subscripts ``c'' and ``s'' indicate
charge and spin fields respectively. 
This relation shows that the operator $\cos \phi_s$ plays the role of the 
amplitude of the order parameter, whereas the operator $\theta_c$ represents
the superconducting phase. 
We have proved \cite{ekz} that when pairs of holes hop on and off a stripe, 
they acquire a gap in their spin degrees of freedom because the undoped 
regions have a spin gap or pseudogap. As a result, $\cos \phi_s$ 
acquires a finite expectation value, the amplitude of the superconducting
order parameter becomes well-defined, and local quasi 
one-dimensional superconducting fluctuations become significant. This takes
place at the lower crossover temperature $T^*_2$, which is essentially a 
property of a single stripe and so is relatively insensitive to the value of 
$x$, until it is cut off by $T^*_1$ at larger dopant concentrations.  
Throughout the underdoped regime
$T_c$ is determined by pair hopping between stripes, and 
ultimately between the planes. This process is the microscopic version of the
phase coupling in Eq. (1).

The order parameter also acquires its symmetry from the spin degrees of 
freedom. Nematic order breaks fourfold rotational symmetry and would lead 
to a mixed $s$ and $d$-wave symmetry of the superconducting order parameter, 
even in an otherwise tetragonal material.

\subsection{The Relation between Spin and Charge}

The topological nature of the stripes \cite{topo} indicates a strong
correlation of the spin and charge collective modes that is well supported
by the neutron scattering experiments. \cite{jtran,mook}
At the same time, on an intermediate length scale, there is a separation
of spin and charge on an individual stripe, as in a 1DEG. \cite{ekz} This
dual relation between spin and charge is characteristic
of a doped insulator in two dimensions, and it is of central importance for 
overcoming the Coulomb problem. The point is that
A) pairing has its origin in insulating regions of the material, where
the energetic cost of having localized holes in Cu $3d$ orbitals has been 
paid in the formation of the material, and B) on a stripe, the objects that form 
pairs are neutral fermions (spinons), which are not impeded by the Coulomb 
interaction. The validity of this picture is based on the well-established
techniques developed in the theory of the 1DEG.\cite{ekz}

{\bf Acknowledgements:}  We would like to aknowledge frequent discussions of
the physics of {\hts} with J.~Tranquada and to thank M.~Wallin, K.~Moon, and 
S.~M.~Girvin for permission to use Fig. 2.
This work was supported at UCLA by the National Science Foundation grant 
number DMR93-12606 and, at Brookhaven,  by the Division of Materials Sciences,
U. S. Department of Energy under contract No. DE-AC02-76CH00016.


\begin{references}


\bibitem{BM} J.~G.~Bednorz and K.~A.~M\"uller, Z. Phys. B {\bf 64}, 189 (1986).

\bibitem{arpes} Z.-X.~Shen {\it et al.}, {\it Science} {\bf 267}, 343 (1995).

\bibitem{muon} Y.~J.~Uemura {\it et al.}, {\prl} {\bf 62}, 2317 (1989);
{\prl} {\bf 66}, 2665 (1991).

\bibitem{optical} B.~Batlogg in  {\it High Temperature Superconductivity},
edited by K.~S.~Bedell {\it et al.} (Addison-Wesley, Redwood City, 1990),
p. 37. 

\bibitem{batloggemery}  B.~Batlogg and V.~J.~Emery, {\it Nature}
{\bf 382}, 20 (1996).

\bibitem{ekz}  V.~J.~Emery, S.~A.~Kivelson, and O.~Zachar, 
{\prb} {\bf 56}, 6120 (1997).

\bibitem{ekl} V.~J.~Emery, S.~A.~Kivelson, and H.-Q.~Lin, 
{\prl} {\bf 64}, 475 (1990). 

\bibitem{hellberg} 
C.~S.~Hellberg, J. Phys. Chem. Solids, this conference.

\bibitem{ute} U.~L\"ow {\it et al.},
Phys. Rev. Lett.  {\bf 72}, 1918  (1994).


\bibitem{spherical} L.~Chayes {\it et al.}, 
Physica A{\bf 225}, 129 (1996).  

\bibitem{zaanen} J.~Zaanen, J. Phys. Chem. Solids, this conference.

\bibitem{jtran} J.~M.~Tranquada, {\it Proceedings of the International
Conference on Neutron Scattering} Toronto, Canada (1997), to appear in 
Physica {\bf B}. (cond-mat/9709325)


\bibitem{mook} H.~A.~Mook,  J. Phys. Chem. Solids, this conference.

\bibitem{cheong} S-W.~Cheong,  J. Phys. Chem. Solids, this conference.

\bibitem{physica} V.~J.~Emery and S.~A.~Kivelson, Physica C {\bf 209}, 597 (1993).

\bibitem{dicastro} C. Di Castro,  J. Phys. Chem. Solids, this conference.

\bibitem{julien} M.-H.~Julien {\it et al.}, {\prl} {\bf 76}, 4238 (1996).

\bibitem{batlogg} B.~Batlogg {\it et al.}, J. Low Temp. Phys. {\bf 95}, 23 
(1994); Physica {\bf C} 235-240, 130 (1994).

\bibitem{homes}  C.~C.~Homes {\it et al.}, {\prl} {\bf 71}, 1645 (1993).

\bibitem{basov} D.~N.~Basov {\it et al.}, {\prl} {\bf 77}, 4090 (1997).

\bibitem{arpesgap}  A.~G.~Loeser {\it et al.}, {\it Science} {\bf 273}, 325 (1996).
H.~Ding {\it et al.}, {\it Nature} {\bf 382}, 51 (1996).


\bibitem{nature}  V.~J.~Emery and  S.~A.~Kivelson,  
Nature {\bf 374}, 4347 (1995). 

\bibitem{oda}  M.~Oda {\it et al.}, J. Phys. Chem. Solids, this conference.

\bibitem{hardy} W.~N.~Hardy, {\it et al.} in {\it Proceedings of the 10th Anniversary 
HTS Workshop}, edited by B.~Batlogg {\it et al.}, 
(World Scientific, Singapore (1996) p. 223.

\bibitem{quantum}  S.~Chakravarty, V.~J.~Emery, and  S.~A.~Kivelson, 
unpublished.  

\bibitem{dimitri} K.~Zhang {\it et al.}, {\prl} {\bf 73}, 2484 (1994);
D.~N.~Basov {\it et al.}, {\prl} {\bf 74}, 598 (1995).


\bibitem{figure} M.~Wallin, K.~Moon, and S.~M.~Girvin, unpublished.

\bibitem{quasi} P.~J.~Hirschfeld and N.~Goldenfeld, {\prb} {\bf 48}, 4219 (1993);
P.~J.~Hirschfeld {\it et al.}, {\prb} {\bf 50}, 10250 (1994);
D.~Xu {\it et al.} {\prb} {\bf 51}, 16233 (1995);   
W.~Atkinson {\it et al.}, Turkish J. Phys. {\bf 20}, 670 (1996).

\bibitem{lee}   P.~A.~Lee, J. Phys. Chem. Solids, this conference.

\bibitem{millis} A.~J.~Millis, J. Phys. Chem. Solids this conference;
A.~J.~Millis {\it et al.}, preprint (cond-mat/9709222). 

\bibitem{ding} H.~Ding {\it et al.}, {\prl} {\bf 78}, 2628 (1997).

\bibitem{stripesqp} This is clear in the stripe model of doped insulators.

\bibitem{gil} V.~B.~Geshkenbein {\it et al.} {\prb} {\bf 55}, 3173 (1997).

\bibitem{kfe}  S.~A.~Kivelson, E. Fradkin, and V.~J.~Emery, to be published.

\bibitem{djs} D.~J.~Scalapino,  J. Phys. Chem. Solids, this conference.

\bibitem{kondo}  
V.~J.~Emery and S.~A.~Kivelson, 
J. Phys. Chem. Solids, {\bf 53}, 1499 (1992). 

\bibitem{varma} C.~M.~Varma,  J. Phys. Chem. Solids, this conference. 

\bibitem{darpes} Z-X.~Shen {\it et al.}, {\prl} {\bf 70}, 1553 (1993);
H.~Ding {\it et al}, {\prb} {\bf 54}, R9678 (1996).

\bibitem{topo} S.~A.~Kivelson and V.~J.~Emery, Synthetic metals, {\bf 80},
151 (1996).

\end{references}
\end{document}